\documentclass[aps,prl,amsmath,twocolumn]{revtex4}
\usepackage{bm}
\usepackage{graphicx}

\begin{document}

\title{ From Burgers to Navier-Stokes  turbulence}

\author{ K.P. Zybin$^*$,
}

\affiliation{ 119991 Theoretical Department of P.N.Lebedev Physical Institute, Russian Academy of Sciences, Moscow, Russian Federation}

\affiliation{101000 National Research University Higher School of Economics, Myasnitskaya 20, Moscow, Russian Federation}

\begin{abstract}
It is shown that the origin of the Kolmogorov's law of the fully developed  turbulence is the result of the joint  stochastic dynamics of pair points separated by the shock. The result obtained in 1-d case generalized on 3-d turbulence. A novel procedure of determination  of correlation functions in 3-d turbulence is proposed.   

\vspace{0.4cm} \noindent {\small PACS numbers: 47.10.ad, 47.26.E-, 05.40.-a}
\end{abstract}

\maketitle

Burgers equation was scrutinized  (see \cite{beck}) because of many physical applications. I will restrict my consideration to the problems of fluid turbulence only. Originally Burgers equation was introduced as a simplification of the Navier-Stokes (NS) equation with
 hope of shedding some light on issues such as turbulence. Besides,  there is still an idea that  time evolution of NS equation  leads to singularity (in the limit of zero viscosity) which is responsible for Kolmogorov scaling \cite{frisch},\cite{kuzn}. It is whell known that without viscosity the time evolution of  Burgers equation results in singularity $v\propto x^{1/3}$ but it exists for a moment only and it is not connected with the velocity statistics.

Stochastic   Burgers equation under the action of large-scale force takes a form
\begin{equation}\label{BE}
 v_t + v v_x = \nu v_{xx} + g(x,t)
\end{equation}
Large scale forcing $g(x,t)$ implies the existence of the inertial interval $\eta(\nu) \ll l \ll L$
analogous to NS equation, where $\eta(\nu)$ is a dissipative scale and $L$ is the scale of the force. Actually, the
structure functions of velocity increments obtained from numerical simulation of Burgers equation exhibit scaling law analogous to the real turbulence \cite{beck}:
\begin{equation}\label{sbur}
S_n=\left<[v(x+l,t) - v(x,t)]^n\right> \propto l^{\xi_n}
\end{equation}
when $\xi_n = n/3$ for $n < 3$ and saturated at the level $\xi_n = 1$ for $n\geq3$. The saturation of the scaling exponents could be naturally understood in the framework of correlations of the step functions. As is well known the long time evolution of the equation (\ref{BE}) results in the appearance of shocks (in the limit $\nu \to 0$). One can easily find that correlation of the shocks gives the following scaling law:
$$
S_n(shock) = \left< [sign(x+l) - sign(x)]^n\right> \propto l
$$
As for structure functions exponents law $\xi_n = n/3$, there is dimensional analysis \cite{yakh}, and exact result for the 3-d order structure function, plus Polyakov's idea of spontaneous breaking of Galilean invariance and dissipative anomaly \cite{polak}, but there is no analytical calculation of them.   Polyakov's approach was based on introduction of the characteristic function of  $n$-point velocity distribution
\begin{equation}\label{Zn}
Z_n(\lambda_j,x_j,t) = \left<exp\left\{\sum \lambda_j v_j(x_j) \right\}\right>
\end{equation}
Structure functions could be obtained from (\ref{Zn}) by differentiating on $\lambda_i$ and  then putting the lambdas zero :
$$
\left<v_{i_1}....v_{i_k}\right> =
\left. \left(\partial^k Z_n\right) / \left(\partial \lambda_{i_1} ... \partial \lambda_{{i_k}}\right)\right|_{\lambda_i = 0}
$$
This approach was generalized in  case of generating functional   for velocities and  velocity derivations \cite{migdal}. But this quantum field theory consideration allows to find some instanton solution when $\lambda$ is a big parameter \cite{polak}-\cite{migdal}.
As a result the obtained solution does not allow to get structure functions by differentiating this relation on $\lambda$ and putting $\lambda = 0$ afterwords. Thus the statistical description of Burgers turbulence does not exists.

The aim of my paper to develop statistical theory   for Burgers equation and determine   the structure functions exponents in the case of NS equation.

For our purpose it is more convenient to use another function $F_N$ which is Laplas transformation of (\ref{Zn}):
$$
F_N = \left<\theta (u_1 - v_1(x_1)) \theta(u_2 - v_2(x_2))\, ...\,\theta(u_n - v_n(x_n))\right>
$$ 
Probability distribution function could be easily determined from this relation by the formula:
$$
P_n  = \left(\partial^n F_N\right) / \left(\partial u_1 ... \partial u_n\right)
$$

One can get  for $F_N$  an equation \cite{polak}:
\begin{equation}\label{eq_FN}
\partial_t F_N + \sum{u_k\partial_{x_k} F_N} - \sum{\kappa(x_i-x_j)\partial^2_{u_iu_j} F_N} =
\end{equation}
$$
D_N =\sum \partial_{u_i}\Gamma^{i}_N
$$
here  $\Gamma^{i}_N$ takes the form:
$$
\Gamma^{i}_N = \nu\partial^2_{yy} \int u_0\,du_0 \partial_{u_0}F_{N+1}\left(u_0, x_i+y; u_1, x_1; ... u_N, x_N\right)
$$
The equation (\ref{eq_FN}) is obtained directly from (\ref{BE}) in a proposition of Gaussiatity of the large scale force $g(x,t)$:
$$
\left<g(x_1,t_1)g(x_2,t_2)\right> = \kappa (x_1 - x_2) \delta (t_1 - t_2)
$$
Let us consider equation for $F_2$ function, the term $D_2$ could be presented in the form:
$$
D_2 = lim_{x_0\to x_1} \left(\partial_{u_1}d_1 + \partial_{u_2}d_2\right)
$$
here 
$$
d_1 = \nu \partial^2_{x_0 x_1}
\int u_0 \, du_0\partial_{u_0} F_3(u_0,x_0;u_1,x_1;u_2,x_2)
$$
The term $d_2$ could be written analogously. It is necessary to remember, that function $F_3$ here is
$$
F_3 = \left<\theta(u_0-u_1(x_0))\theta(u_1-u_1(x_1))\theta(u_2-u_2(x_2))\right>
$$
hence
$$
\partial_{x_0}\int u_0 du_0\partial_{u_0} F_3 = \left<\partial_{x_0} u_1\theta(u_1-u_1(x_1))\theta(u_2-u_2(x_2))\right>
$$
To calculate derivation on $x_1$ it is necessary to differentiate on $u_1$ before. So we have:
\begin{equation}\label{d1}
d_1 = \nu \left<\partial_{x_0} u_1 \partial_{x_1} u_1\theta(u_1-u_1(x_1))\theta(u_2-u_2(x_2))\right>
\end{equation}
To calculate it in the limit $\nu\to 0$ it is necessary to know the solution of (\ref{BE}) at ultraviolet limit. But we know that it  becomes a step function. Actually for smooth  large-scale force $g(x,t)$ let us solve the equation (\ref{BE}) by standard methods of expansion of variables and matching obtained asymptotic solutions \cite{naif}. Introducing a new variable $y = x/\nu$ one can get from (\ref{BE}):
\begin{equation}\label{Bnu}
\partial^2_{yy} v - v \partial_yv = \nu \left( \partial_t v - g(\nu y,t)\right)
\end{equation}
Now let us fix the variable $y$ and put $\nu = 0$. The main asymptotic takes a form
$$
2 \partial_y v = v^2 - v_0^2
$$
After integration of this well-known equation  we will get
\begin{equation}\label{stupen}
v = v_0 \left(1 - e^{v_0x/\nu}\right)/\left(1 + e^{v_0x/\nu}\right) + U_0
\end{equation}
We choose $v=U_0$ if $y=0$.
As we  see it is enough to take into account antisymmetric part of the velocity only, the symmetric part is  smooth  and does not give any input into dissipation.

To match this solution to an external one let us consider $F_1$ function.
$$
\kappa(0) \partial^2_{u_1u_1} F_1 + D_1 = 0
$$
Multiplying this equation on  $u_1^2$ and integrating over $du_1$ we find:
$$
\kappa(0) = \nu\left. \left<\partial_{x_0} u_1 \partial_{x_1} u_1\right>\right|_{x_1\to x_0}
$$
Now let us {\it suppose an ergodicity} of the obtained solutions. It is a usual proposition in experimental study of the developed turbulence. On the basis of this proposition and taking into account (\ref{stupen}) we get:
\begin{equation}\label{dv2}
\left<\nu \left(\partial_x v\right)^2\right> = v_0^3/(3 L)
\end{equation}
This relation should be fulfilled in the limit $L\to\infty$. Matching it with the previous one we get an idea of ``infrared anomaly'':
\begin{equation}\label{UV}
lim_{L\to\infty} \left[v_0^3(L)/(3 L)\right] = \kappa(0)
\end{equation}
Now let us return  to expression for $d_1$ (\ref{d1}).  In accordance with ergodic hypothesis we define average as:
$$
<A(x_1)\theta[x_1]\theta[x_2]> = \frac{1}{2L}\int_{-L}^{+L} A(x_1+y)\theta[x_1+y]\theta[x_2+y]\, dy
$$
Here $\theta[x_1]$ is a symbolic name of theta function from $F_N$ definition. According to this definition the average (\ref{d1}) could be rewritten as a space integral.

As it was mentioned above, our distribution function $F_2$ contains either fast  or slow variables and solution (because of linearity)  could be presented as a sum $f_2(fast) + f_2(slow)$. It is very easy to find $f_2(fast)$. In fact combining it from solution (\ref{stupen}) one can get:
$$
f_2(fast) = \frac{1}{2L}\int_{-L}^{+L} \theta(u_1-v_1(x_1 - x_2 +y))\theta(u_2-v_2(y))\,dy 
$$
Where $v_{1,2}(x)$ are smooth-step-like solution (\ref{stupen}). Analogously one can introduce $f_n(fast)$ for arbitrary $n$. Obviously it will be a solution for generating function (if we neglect an interaction between steps) since (\ref{stupen}) is a solution of the Burgers equation and diffusion on velocitied does not give any input into function $f_2(u_1-u_2, x_1-x_2)$. 
In the limit $\nu\to 0$ this function becomes an average set of steps and naturally gives the scaling $<(\delta v)^n> \propto l$ for $n>1$.
Besides, in this limit the probability of the coincidence of two shocks in one point is equal zero. 

Now let us separate the slow part of the distribution function. If theta functions changes slowly on the scale $1/\nu$ the production $\partial_1 u \partial_2 u$ could be presented (in the limit  $\nu\to 0$)as a delta function (see (\ref{dv2})):
$$
lim_{\nu\to 0} \left<\partial_{x_0} u_1 \partial_{x_1} u_1\right> = \left[v_0^3(L)/(3 L)\right] \delta \left(x_0 - x_1 \right)
$$
Thus $d_1= v_0^3/(3L)$
$$
\times \int dy \delta (x_0-x_1-y) \partial_{u_1}\theta(u_1-u_1(x_1+y))\theta(u_2-u_2(x_2+y)) 
$$
$ \approx [v_0^3/(3L)] \partial_{u_1} f_2(u_1,x_1; u_2,x_2) $

Taking into account (\ref{UV}), we will find:
$$
u_1\partial_{x_1} f_2 + u_2\partial_{x_2} f_2 =\sum_{i,j=1,2}{\kappa(x_i-x_j)\partial^2_{u_iu_j}f_2} +
$$
$$
 \kappa(0)\left[\partial^2_{u_1u_1} f_2 + \partial^2_{u_2u_2} f_2\right]
$$
Due to uniformity condition the distribution function depends on  $l = x_1-x_2$ only. Introducing  $v=u_1-u_2$ and $U=u_1+u_2$, after integration over $dU$ we have got:
\begin{equation}\label{fin}
v\partial_l f =[2\kappa(0)-\kappa(l)-\kappa(-l) + \kappa(0)] \partial^2_{vv}f
\end{equation}
Taking into account that $l << L$, the equation takes a simple form:
$$
v\partial_l f = \kappa(0) \partial^2_{vv}f
$$
It is important to note that stationary solution of this equation exists only if $(v l) > 0$. That is why structure functions $<|v|^n>$ are not well defined. In numerical modeling $<|v|^3>\propto l^{0.85}$ but is not coincide with exact analytical result  $l^1$  \cite{085}.  Possibly the absence of stationarity for $(v l) < 0$ is a reason for the difference.

Let us make  Laplas transformation of the equation (\ref{fin}) and consider the limit $k\to\infty$, which corresponds to $l\to 0$. In this case one can find WKB solution:
$$
f \propto exp\,\left\{ -2 \sqrt{k} v^{3/2}/3\right\}\,,\qquad\qquad k\approx [l\,\kappa(0)]^{-1}
$$
It is easy to get  Kolmogorov's law by integrating this distribution function:
$$
\left< v^n\right> = \int v^n f\, dv\,\propto l^{n/3}
$$

Thus we have combination of the solutions:
$$
\left< v^n\right> = C_1\, l  + C_2 \, l^{n/3}
$$
It means that for $0 < n < 3$ in the limit $l\to 0$ we have Kolmogorov's law  $n/3$ and after $n>3$ we have saturation caused by shocks correlation.

To understand this result more clearly let us derived the equation (\ref{fin}) directly from expansion of (\ref{Bnu}) on parameter $\nu$. First let's put the typical scale  of the external force equal to infinity and $g(+\infty) = g_+$ correspondingly $g(-\infty) = g_-$. 

The condition of the existence of the $\nu$ expansion is the absence of secular term. Integrating (\ref{Bnu}) over $dy$ from $-R_1\to -\infty$ to $R_2\to +\infty$ on can get:
$$
\int_{-R_1}^{R_2} \left[2\partial_t v - (g_+ + g_{-})-(g_+ - g_{-})\,sign\,y\right] dy < \infty
$$
this relation and (\ref{stupen}) gives:
$$
2\partial_tU_0 = (g_+ + g_{-}) =\phi(t) \,,\qquad 2\partial_tv_0 = (g_+ - g_{-}) = \psi(t)
$$
It is necessary to note also that we should  choose $v_0=0$ if $\psi = 0$ and in this case there is no shock and as the result there is no dissipation on it.

Now let us take a look at the dynamics of two points $x_1(t)\,,x_2(t)$. If these points are both located on the left or on the right side of the shock the velocity difference $v=v_1(t) - v_2(t)$ and distance $l = x_1(t) - x_2(t)$ does not change with time. 

But if we consider a case when,  $x_1 > 0$ and $x_2 <0$ or opposite, the distance between these particles $l$ and velocity difference $v$   changes with time:
$$
\partial_t l = v \,,\qquad \qquad \partial_t v = \psi
$$
One can introduce probability distribution function
$$
P(v,l,t) = \left<\delta(l - l(t))\delta(v-v(t))\right>
$$
Considering  $\psi(t)$ as Gaussian stochastic process it is easy to get equation
$$
\partial_t P + v \partial_l P = D \partial^2_{vv} P
$$
which in stationary conditions coincides with (\ref{fin}). So we see from our consideration that Kolmogorov's input into structure function is connected with joint stochastic dynamics of the points separated by the shock. It is worth  noting that this equation is equivalent to infinite chain for the structure functions $S_n = <v^n>$. Actually multiplying (\ref{fin}) by $v^{n}$ one can get:
\begin{equation}\label{chain}
\partial_l S_{n+1} = D S_{n-2}
\end{equation}
It is easy to see that $S_n\propto l^{n/3}$ is the solution of the equation (\ref{chain}).

Thus we see that according to discussed theory the structure function exponents should be bi-fractal if you calculate them in the region $v_il_i >0$.

Let us try to apply the ideas discussed above to the NS equation. The success of the above approach is based on the fact that we know singular-like solution of the Burgers equation at $t\to\infty$. In the case of NS equation a singular-like solution  was discussed in \cite{ZS_PRE},\cite{ZS_UFN}.  It was obtained  in inertial interval by introducing large-scale velocity $U_i(x,t)$ instead force. The external force was determined by equation:
$$
\partial_t U_i + U_j\partial x_j U_i = F_i(x,t)
$$
It is necessary to emphasize that $U_i$ is given function and this  equation is the definition of the force $F_i$ in the inertial interval. But inside inertial interval force $F_i$ does not work. All the energy input is defined by large-scale forcing. In this case one can choose $U_i$ as uniform and isotropic   Gauss stochastic process (or arbitrary symmetric in time process). Actually  in this case we have $<U_i F_i> = 0$. Because of large scale natura of the field $U_i$ it is possible to expand it into Taylor series. The first term is the most important if we restrict our self  by relation $l << L$.

It was shown that under the action of large-scale velocity gradient tensor $A_{ij}(t)$
in the solution of NS equation appears (in the limit $t\to\infty$) singular-like vorticity $\omega$   \cite{ZS_PRE},\cite{ZS_UFN}. Such kind of solution arises with unit probability.  The solution looks like a ribbon, where  $\omega$ is directed along the ribbon (see Fig.1). The vorticity is concentrated exponentially in time in a sheet which has stochastic rotations. In a local frame we have the  strongest expansion along $\omega$ and contraction on perpendicular direction. An account of viscosity gives quasi stationary solution. 
\begin{figure}
\includegraphics[width=8cm]{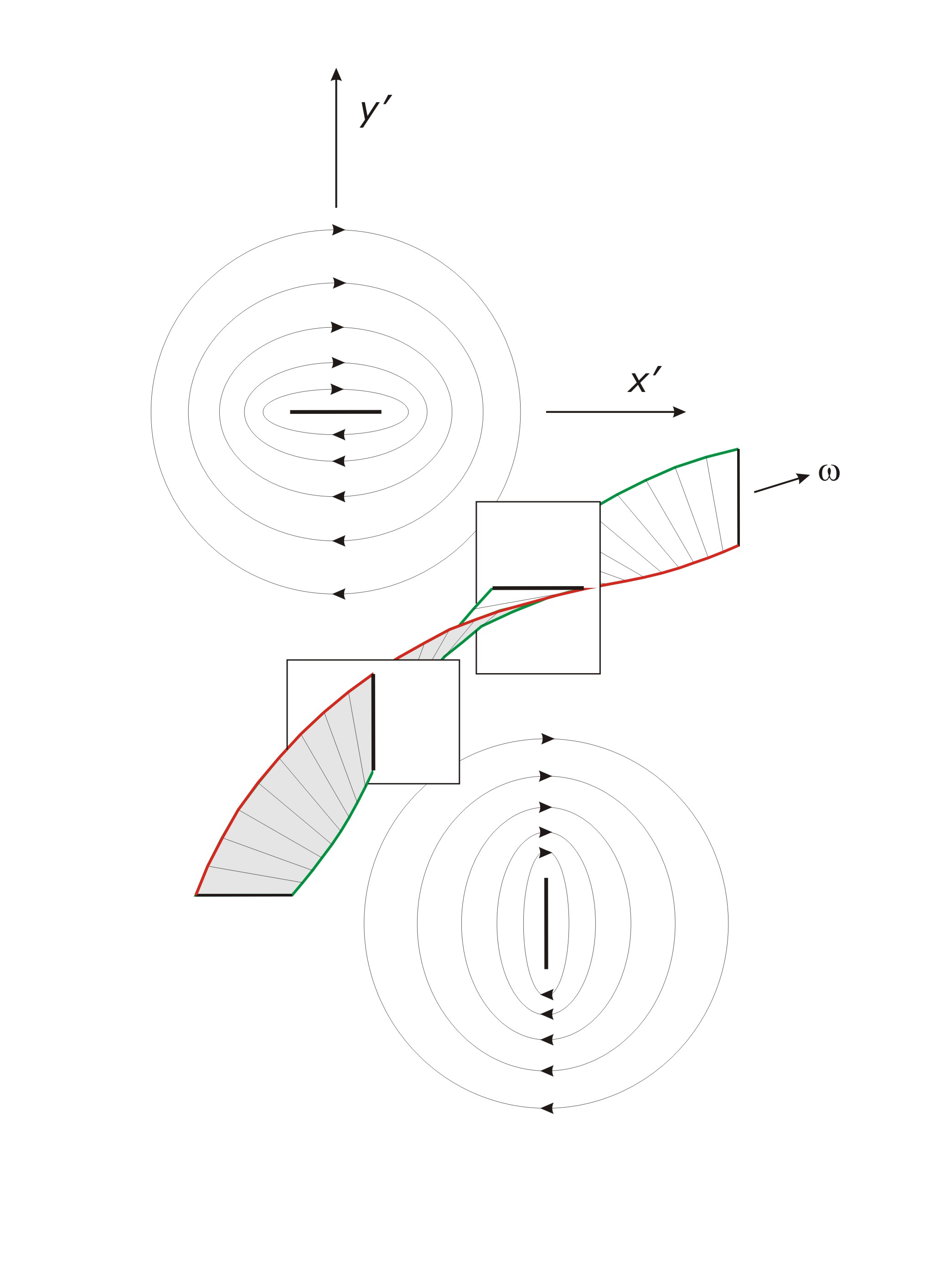}
\caption{ribbon-like solution, the vorticity concentrated in the central layer (ribbon),  circus with arrows show velocity}
\end{figure}
Due to quasi one-dimension character of the solution (in general case we have contraction along one direction) the viscosity is important along this direction  only. The stationary solution in the rest frame takes the form \cite{ZS_PRE}:
\begin{equation}\label{NSstep}
\lambda_1x'\partial_{x'}\omega - \lambda_3\omega = \nu \partial^2_{x'x'}\omega
\end{equation} 
Here parameters $\lambda_1$ and $\lambda_3$ are Lapunov's exponents responsible for the discussed above local vorticity dynamics. To define them it is necessary to solve stochastic matrix equation:
$$
\partial_tQ_{ij} = Q_{ik}A_{kj}(t) 
$$
According to a set of theorems (see for the details \cite{ZS_PRE},\cite{ZSIlin}) the matrix $Q$ in the limit $t\to\infty$ has (with unit probability) solution:
$$
Q =  Z_{\infty} e^{D t} R(t)\,,\qquad \qquad D(t) = diag\left\{ \lambda_1\,,\lambda_2\,,\lambda_3\right\}
$$
here $Z_{\infty}$ is a number matrix and $R(t)$ is a rotation. Due to incompressibility $\lambda_1 +\lambda_2 +\lambda_3 =0$ where $\lambda_3$  was chosen to be a maximal one.

For any symmetrical in time process the value $\lambda_2 =0$ \cite{ZIlin}. As is known time asymmetry in NS equation is connected with energy dissipation which is determined by large-scale stochastic force but not large-scale velocity gradient tensor. Thus $\lambda_2\approx 0$. Taking this fact into account one can solve the equation (\ref{NSstep}) for vorticity and velocity can be determined by integrating. As a result in the limit $\nu \to 0$ we have (neglecting slow logarithmic relation) a step-like solution analogous to (\ref{stupen}):
\begin{center}
$v_{y'}(x') = v_0 \,sign\, x'$
\end{center}
The main difference of this solution from the solution of Burgers equation is incompressibility. We have got shear solution. It significant however, that it is a kind of ``force free'' solution $U_y(x)$ where pressure has a large-scale gradient along axis $y, z$ but not important along  velocity gradient $x$. 

This step-like solution is obtained in rotation reference frame. Returning  to the fix frame we find:
$v_x = - v_{y'}\,sin\,\phi \,,\quad v_y = v_{y'}\,cos\,\phi\,\quad x'= x\,cos\,\phi + y\, sin\,\phi$

Now let us construct a perpendicular velocity increment $\delta v_{\perp}= v_{\perp}(r) - v_{\perp}(r - l)$. Here $v_{\perp}= v_y cos\,\phi - v_x sin\,\phi$ and $r = x cos\,\phi + y sin\,\phi$. Averaging on rotation $\phi$ and on space, on can get perpendicular structure function:
$$
\left<\delta v_{\perp}^n\right>  = \frac{1}{L^2}\int_0^L\left[ sign\,r - sign\, (r-l)\right]^n r dr   \propto l^2
$$
This solution is analogous to Burgers step solution (\ref{sbur}) and connected with fast part of PDF or generating function. An equation for generating function for the case of NS  equation was obtained in \cite{yak_chain} and it is quite analogous to equation (\ref{eq_FN}). For our further purpose we will consider equivalent presentation of generating function -- infinite chain of structure function equations (for details see \cite{hill}):
$$
\partial_t D_N + \nabla_r D_{N+1} = - T_N + 2\nu \left[ \nabla_r^2 D_N - E_N\right] 
$$
here $D_n = <v_j... v_i>\,,\qquad E_N =<v_k....v_l e_{ij}>$

and $ T_N =<v_j....v_l(\partial_{x_i}p - \partial_{x'_i}p')> $

While averaging $<\,>$ the sum of all terms of a given type that
produce symmetry under interchange of each pair of indices were taking into account. And $[N]$ within square brackets denote the number of indices.

As in the case of Burgers equation there is a very important value $e_{ij}$ connected with energy flux:
\begin{center}
$e_{ij} = \partial_{x_n}u_i \partial_{x_n}u_j + \partial_{x_n'}u_i'\partial_{x_n'}u_j'$
\end{center}
Actually, (see \cite{hill})  $2\nu E_{11} = 2\nu E_{22} = 4\varepsilon/3$  where  $\varepsilon$ is the energy flux. But this value changes greatly on the dissipation scale. Following ideas of solution (\ref{fin}) let us  restrict  our consideration to a slow-changing part of the distribution function. It means that in calculation of $E_N$ all the  values except $e_{ij}$ changes slowly (just like in the case of Burgers equation). Thus
\begin{equation}\label{EN}
E_N =< \left\{v_k....v_l e_{ij}\right\}> = C_0 \varepsilon D_{N-2}
\end{equation}
Here $C_0$ is a constant. Actually, this term could be rewritten in the terms of distribution function $F_N(v_1,..v_N; x_1,..x_N)$ as:
$$
E_N =\int e_{ij}(r,t)\, v_k...v_l\, F_N \,Dv 
$$
and if the distribution function changes slowly in space and time we have the expression (\ref{EN}).

 We  are going to find the solution of the chain in the limit $l\to 0$ (we have already eliminate $\nu$ introducing $\varepsilon$ and can put $\nu = 0$ now). But according to our ribbon - like solution  in the limit $l\to 0$ pressure does not play any role because it is a shear singular-like solution.  The pressure gives input on the scale of the order  of ribbon's radius and could be neglected in our consideration.  As a result the slowly changing  part of the chain in the limit $\nu\to 0$ takes a form: 
\begin{equation}\label{kolmsol}
 \nabla_r D_{N+1} = C_0\varepsilon  D_{N-2}
\end{equation}
One can see that there is Kolmogorov's  scaling solution to this chain:
$D_{N}\propto r^{N/3}$

Now let us discuss the  results obtained. We see that our solution of the NS equation is very close to the solution of the Burgers equation (\ref{sbur}). Actually, structure functions $S_n^{\parallel\perp}$ obey  Kolmogorov's law for $0 < n < 6$ and after that we have saturation solution $S_n^{\perp} \propto l^2$ but for perpendicular structure functions only; ribbon-like solution does not give input into main asymptotic of longitudinal structure functions. According to modern numerical simulations \cite{bens},\cite{got} the value of  perpendicular structure function exponents is less than longitudinal one and more close to Kolmogorov's law. This fact agrees with  singular-like structures discussed above. 

The obtained equation (\ref{kolmsol}) is analogous to (\ref{chain}). They seem to  have close physical interpretation. In  case of NS equation the dynamics of pair points,  separated by step, is responsible for Kolmogorov's law too. But in  case of NS equation we have step of tangential velocity.

The similarity between  NS and Burgers stochastic solution could be seen also in equations for PDF. In the case of NS, the stationary equation for  PDF is a diffusion type too, and the left side of the equation is $v_i\partial_{l_i} f_2$. This term should be positive.

Thus, in both case -- Burgers and NS turbulence theory predict bifractal behavior of structure functions if condition for velocity difference $v_il_i > 0$ is taken into account. 

Author is grateful to A.S.Il'yn and M.O.Ptitsyn for discussion.

\end{document}